\documentclass[sigconf]{acmart}

\pdfoutput=1
\usepackage{cleveref}
\crefname{section}{}{\S\S}
\crefdefaultlabelformat{\S#2#1#3}
\usepackage[font=footnotesize]{caption}

\AtBeginDocument{%
  \providecommand\BibTeX{{%
    \normalfont B\kern-0.5em{\scshape i\kern-0.25em b}\kern-0.8em\TeX}}}

\begin{document}

\title{Simulating Performance of ML Systems with Offline Profiling}

\author{Hongming Huang$^\ast$, Peng Cheng$^\dagger$, Hong Xu$^\ast$, Yongqiang Xiong$^\dagger$}
%%\authornote{Both authors contributed equally to this research.}
% \email{honhuang7-c@my.cityu.edu.hk}
%%\author{Possible Equal Author}
%%\authornotemark[1]
%%\email{foo@bar.com}
\affiliation{%
 \institution{$^\ast$Department of Computer Science, City University of Hong Kong}
 \state{$^\dagger$Microsoft Research}
 % \streetaddress{Tat Chee Avenue, Kowloon}
 % \city{Hong Kong}
 %%\state{Hong Kong}
 % \country{}
}
\begin{comment}
\author{Peng Cheng}
\email{pengc@microsoft.com}
\affiliation{%
 \institution{Microsoft Research Asia}
 \streetaddress{Danling Strict, Haidian,}
 % \city{Beijing}
 %%\state{Hong Kong}
 % \country{China}
 }

\author{Hong Xu}
\email{henry.xu@cityu.edu.hk}
\affiliation{%
 \institution{City University of Hong Kong}
 \streetaddress{Tat Chee Avenue, Kowloon}
 % \city{Hong Kong}
 %%\state{Hong Kong}
 % \country{China}
 }
\end{comment}

\begin{abstract}
  % State-of-art distributed machine learning system can explore optimal
  % distributing strategy by online profiling of ML system. However, this
  % approach is expensive and limited by hardware. 
  We advocate that simulation based on offline profiling is a promising approach
  to better understand and improve the complex ML systems. 
  Our approach uses operation-level profiling and dataflow based simulation to
  ensure it offers a unified and automated solution for all frameworks and ML
  models, and is also accurate by considering the various parallelization
  strategies in a real system.
  % \\ 
  % Our preliminary results shows that offline
  % profiled operation performance is stable (<1\% variance)and predictable and
  % simulator on real model is also accurate (<2\% simulation error), which
  % proved the feasibility of our approach.
\end{abstract}



% \keywords{distributed machine learning, system, simulation}

\acmConference[MLOps'20]{ACM Conference}{Mar 2020}{Austin,
	TX, USA}
\maketitle

%!TEX root = main.tex

\section{Introduction}
\label{sec:intro}

% Deep neural networks (DNNs) have been widely adopted to solve many practical
% problems, such as image classification, neutral language processing, etc. 

% performance profiling very important in distributed ML sys. existing method is
% online end-to-end profiling. good thing is it's accurate
Machine learning (ML) systems are prevalently used to train the complicated
deep neural network (DNNs)
\cite{chelba2013one,deng2009imagenet,huang2017densely,szegedy2016rethinking}
using a cluster of machines. In this context, characterizing the
performance of an ML system becomes critical for optimizing training strategy
\cite{krizhevsky2014one,NHPS19,jia2018beyond} and system design
\cite{jia2018exploring,mirhoseini2017device}. % pipedream, flexflow; some paper
% on PS improvement
Intuitively, performance can be obtained by actually training a given DNN
model over a given hardware cluster with a given strategy (e.g. parallelization,
hyperparameters, etc.) and measuring its throughput. This {\em online
profiling} approach is the de facto solution used by most existing work 
\cite{NHPS19,jia2018beyond}.

Online profiling yields the most accurate performance. Yet it has several
fundamental limitations that we believe make it ill-fitted for many practical scenarios.
% online profiling's problems:
% 1. expensive, not scalable
% 2. profiling space limited by hardware I have
% 3. blackbox approach, cannot understand or analyze component's impact

Online profiling is very expensive and does not scale. While it is feasible to
run a few iterations of training for a given setup, the exponential number of
potential training strategies---a combination of hyperparameter setting,
parallelization strategy, synchronization and pipelining method, etc.---makes
it impractical to enumerate all possibilities in order to find the optimal
one. Then if we have a new DNN, we have to do profiling all over
again for this new model. Further, online profiling is inherently limited by
the available hardware resources at our disposal. If we wanted to know the
accurate system performance with a new computing or networking
device,
we would have to acquire this hardware first which takes significant time and
financial investments. Finally, due to the complexity of ML systems, online
profiling is typically done in an end-to-end fashion by treating the entire
execution pipeline as a blackbox. Thus it remains difficult today
to dissect and understand the impact of various aspects of the system (say
computation vs communication) in different settings (GPUs, networking (PCI-e,
NV-Link, RDMA, ...), architecture (PS \cite{li2014scaling}, All-reduce
\cite{awan2016efficient}), ...), let alone how to improve the design.

% our idea: build a simulator to estimate mlsys performance based on offline
% profiling
% can solve all problems above; 

We propose a different approach to address these issues. We rely on {\em
offline profiling} to measure performance of basic {execution units} on
different hardware platforms, and based upon it build a simulator to
accurately characterize the system-level performance without actually training
the DNN. Since we profile the basic execution units on tensors including
computation (e.g. conv2d) and communication (e.g. allreduce) instead of the
entire system, offline profiling is much more scalable and the results can
naturally be reused. Different users can easily contribute their profiling
results on their hardware platforms, thus overcoming the hardware constraint.
More importantly, instead of a blackbox approach, our simulator uses the
detailed dataflow graph produced by all major ML systems to accurately
estimate its performance for a given training strategy.

% benefits: use cases
Offline profiling based simulation has promising potentials to become a
foundation for many tasks in MLOps. It accelerates system design, since we can
quickly identify the performance bottleneck of the complex ML system and
project the potential gain of a certain optimization. It also helps many auto ML
and performance engineering tasks, as systems like PipeDream \cite{NHPS19} and
FlexFlow \cite{jia2018beyond} can use it to rapidly find the optimal
parallelization strategy for any DNN, hardware, and hyperparameter settings
without the high overheads of online profiling.

\section{Design}
\label{sec:design}
% challenges: general for all frameworks and all ops; accurate to take into
% account all the execution opt (overlap, etc.)

Building a performance simulator entails two basic questions: (1) how to
profile the basic execution units of a DNN offline, and (2) how to simulate
the system-wide performance using the profiling results? Answering them is
challenging because we have to achieve three objectives at the same time: (1)
the simulator should be {\em accurate} in order to be useful; (2) it should be
{\em general} in order to cover the major dimensions of ML training: framework,
library, hardware,
training strategies, etc.; (3) it should also be {\em efficient} with minimal
human intervention, preferably completely automated.

In the following we explain our key design decisions in answering the two questions
and how they achieve our design goals. 
\begin{comment}
There are two basic challenges we must attack here. 
First, our simulator has to
be accurate in the sense that it needs to considers the wide range of factors
in executing a training pipeline in an ML system, including the overlap of
computation and communication ops, synchronization barrier across ops or
workers, parallelization or pipelining method, just to name a few.
Second, it has to be general in order to cover the major dimensions of ML
training: framework (TensorFlow, PyTorch, ...), library (NCCL, Horovod, ...), hardware,
ops, training
strategies, etc. 
% Lastly, it should minimize the human effort involved, so that a 
\end{comment}

% 2. op-level profiling, op is the atomic unit in DAG: kernel-level too
%    expensive (cuda), hard to automate; layer-level too coarse,
% 3. online profiling for new ops; less on this
\noindent{\bf Op-level profiling.}
There are two types of basic execution units for a ML system. One is operation
(op) which is the atomic execution unit defined at the framework level (e.g.
TensorFlow \cite{abadi2016tensorflow}). The other is kernel which is defined
at the device level (e.g. CUDA for GPU). Op is a
higher level abstraction and is implemented by one or more kernels. We perform
profiling at the op-level that achieves a good trade-off between accuracy and
overhead. We do not favor kernel-level profiling because it requires precise
knowledge of how each op is implemented in the framework using different
kernels. This can only be obtained by analyzing the framework's source code
which is extremely expensive and hard to automate.% On the other hand, op-level profiling only involves
% running an op in a given framework and is much easier.

Even for a single op, it may have many input arguments that makes it difficult to
numerate all possible combinations. For example, the {\tt Conv2D} op in TensorFlow
has four arguments for input tensor shape, four for the filtering kernel shape,
and several other arguments for other attributes of the
2-D convolution \cite{conv2d}. 
Thus we apply a machine learning approach here to reduce the complexity: for each
input argument we profile a fixed number of values, and use these results to train
a neural network to estimate the op performance. 
% This works in
% most cases as execution time on GPU with massive parallelism is usually a linear or
% quasi-linear function of the input size. 
This is also easy to automate. 

% A limitation of offline profiling is that a DNN may have some ops that our simulator
% has not seen before. 
\begin{comment}
No matter how comprehensive our offline profiling can be, it is entirely possible
that a DNN has some ops our simulator has not seen before. 
In this case, we fall back to {online profiling} and measure the new op's performance
by actually running it. 
\end{comment}
% did not say lazy profiling for now

% 1. DAG-based simulation, blueprint general, similar across frameworks; DAG also
% depicts execution strategy for simulation (no dependency across nodes->overlap)

\noindent{\bf Dataflow based simulation.}
Another key design choice is to use the dataflow graph as the basis of our
simulation. Dataflow graph is widely used by all major ML frameworks \cite{mmDNN}
as
the blueprint of execution and is generated automatically according to the user
training
program. 
It is a directed acyclic graph containing the ops as nodes and input/output data
as edges between the nodes. It also represents various
parallelism optimization and distributed execution strategies across devices and
machines. Every node on the graph can be placed on a different machine, which is
indicated in its ``device'' property, and the
synchronization requirements are modeled by dependencies (i.e. edges) across computation
and communication ops\cite{peng2019generic}. All this
information is embedded in the dataflow graph which can be readily obtained with the
framework's API.

Thus our simulator essentially works by ``replaying'' the training execution
based on the dataflow graph to calculate performance. Each independent device
(CPU, GPU, or communication link) executes in parallel and maintains a
job queue and its finish time. The simulator keeps a global
ready list containing all nodes whose dependencies are fulfilled. The
simulator runs in a loop: (1) It starts all nodes in the ready list by
enqueuing them into their corresponding device's job queues. (2) As soon as
an op is finished on a device (using the profiling results), it updates all
successor nodes’ dependency
counter. If the counter becomes zero, the successor node is added into ready
list.
The system performance is obtained by looking at the finish time of the last
device.

 % This step will handle only one
 % node at a time.

\begin{figure}[t]
  \centering
  \includegraphics[width=0.7\linewidth]{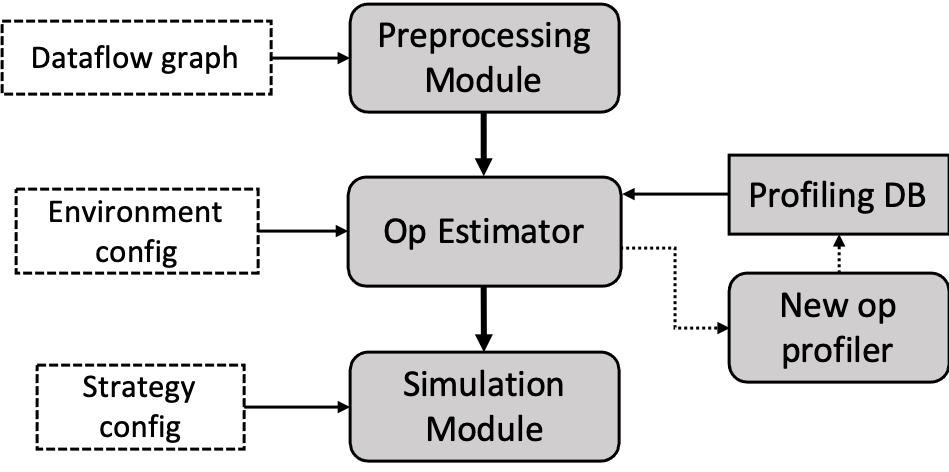}
	\vspace{-4mm}
  \caption{System Overview.}
  \vspace{-7mm}
  \label{fig:overview}
\end{figure}

\noindent{\bf Overall System.}
With these key design choices, our simulator's overall design is shown in 
Figure~\ref{fig:overview}.
% There are 5 modules in the simulator system.
As each framework expresses dataflow graph differently, 
the {\em preprocessing module} first transforms the dataflow graph extracted 
from the framework into a unified format.
The {\em op estimator} estimates performance of each op in the
dataflow graph by querying the {\em profiling database} with offline profiling
results. This requires
information about the training environment (e.g. hardware type,
software library version, etc.) from a config file which is not in the graph.
In case the graph has new ops not in the profiling database, we fall back
to online profiling with the {\em new op profiler} and add the result to the
database.
Then the {\em simulation module} takes the augmented graph, and 
simulates the training execution accordingly and estimates performance as
explained before. 
It also needs additional information about the training strategy from a config
file, such as the number of replicas in data parallelism, and the pipelining
setting for model parallelism which may not be available in the dataflow graph
\cite{NHPS19}.

\section{Preliminary Results}

%\hx{explain the implementation: framework, hardware, lines of code, etc.
%
%for comp, show all the basic ops we have covered. conv2d is just an example
%
%revise the conv2d figure
%
%for comm, explain how you did the profiling
%
%}

% Our experiments seek to answer the following questions:

% 1. Is offline profiling accurate at the op-level?  

% 2. Is the dataflow based simulation accurate in terms of handling the
% parallelism and overlapping in a ML system?

We present our preliminary evaluation results here.

{\em Experiment Setup.} We run experiments on a server with 2 Xeon CPU
E5-2620 and 4 Tesla V100 GPU, running Ubuntu
16.04.5 LTS with CUDA 10.0, cuDNN 7.5, and TensorFlow 1.13.1.

% {\em Key observation.} We have two key observations: 1) Most operations'
% performance has clear relation ship with the input shape and is very stable.
% 2) The simulator algorithm can calculate overlapping and parallelism in real
% training with high accuracy.

\noindent{\bfseries Offline profiling.} We build an automatic profiler with
about 900 LoC in PyThon to profile computation ops. It constructs a dataflow
graph that only contains the input data nodes and 1000 identical
computation nodes corresponding to the op. This is to amortize
the constant overheads of launching the graph and input initialization on GPUs
before training starts. % Then the profiler
% measures the total running time of the graph to obtain the op performance on
% average.
As discussed in \cref{sec:design}, we profile each input argument of the op with
16 possible values.

\begin{figure}[t]
\vspace{-2mm}
	\centering
	\includegraphics[width=0.8\linewidth]{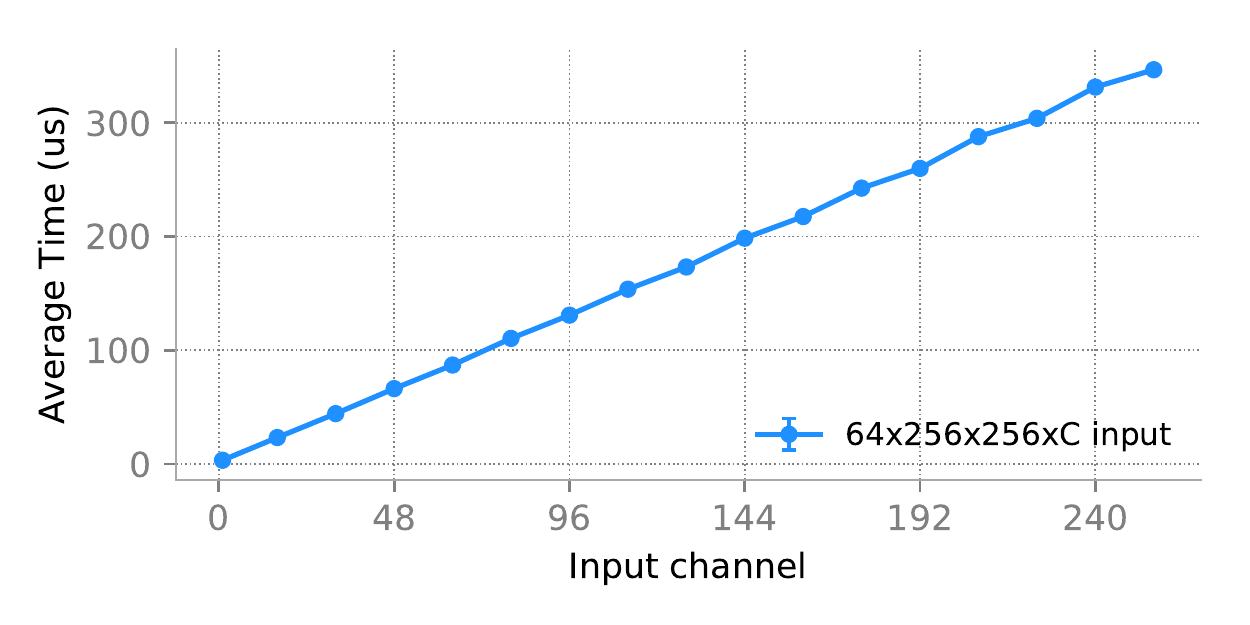}
	\vspace{-5mm}
	\caption{Performance of  {\tt Conv2D} on our testbed.}
	\vspace{-5mm}
	\label{fig:conv2d}
\end{figure}

Our initial experiment profiles over 20 common ops for neural network building
({\tt Relu}, {\tt Sigmoid}, {\tt Conv2DBackpropFilter}, etc.), linear algebra
({\tt MatMul}), and element-wise mathematics. We observe that their
performance is very stable with standard error lower than 1\% of the mean, and
has a strong linear relationship to the input shape. Figure~\ref{fig:conv2d}
depicts the performance of {\tt Conv2D} with varying number of input channels
as an example. This verifies our design choice that we can model and
accurately estimate op-level performance with offline profiling.

% \noindent{\bfseries Communication Profiling.}
We also profile the GPU communication bandwidth in a single machine under
different scenarios. Table~\ref{table:commPerf} shows the result.
% Including: GPU connected via root complex,
% connected via QPI bus, connected via PCI-e switch. 
% Communication performance is quite stable in all these scenarios.
\begin{table}
\footnotesize
	\begin{tabular}{cccl}
		\toprule
		Scenario & QPI & Root Complex & PCI-e Switch\\
		\midrule
		GPU to GPU uni-direct & 10948.81 & 10270.59 & 13181.03\\
		GPU to GPU bi-direct & 16475.93 & 19325.81 & 25037.41\\
		Host to GPU & 11956.69 & 12027.22 & 12347.09\\
		GPU to Host & 13200.21 & 13201.87 & 13121.95\\
		NCCL-AllReduce 2GPUs & 6178.68 & 9162.42 & 11598.12\\
		NCCL-AllReduce 4GPUs & --- & --- & 8048.35\\
		\bottomrule
	\end{tabular}
		\caption{Throughput (MB/s) of GPU communication on a single machine in
		different scenarios.}
	\label{table:commPerf}
	\vspace{-5mm}
\end{table}
\begin{table}[t]
\footnotesize
\vspace{-3mm}
	\begin{tabular}{cccl}
		\toprule
		Model & {\tt TF.timeline} (ms) & Simulation (ms) & Error\\
		\midrule
		VGG\_19 & 203.38 & 199.66 & 1.83\%\\
		ResNet\_50 & 282.56 & 277.47 & 1.80\%\\
		ResNet\_152 & 638.81 & 629.31 & 1.49\%\\
		\bottomrule
	\end{tabular}
		\caption{Per-iteration training time with batchsize 64 on CIFAR-10.}
	\label{tab:simVeri}
	\vspace{-8mm}
\end{table}

\noindent{\bfseries Simulation.}
% on some models, compared to timeline.
We use three common CNN models to evaluate the accuracy of
dataflow-based simulation. 
We use the {\tt TF.timeline} to measure the actual training time. 
Since our profiling is not complete yet, we use the offline profiling results
whenever applicable, and rely on {\tt TF.timeline} to do online profiling for other ops. 

% \begin{table}
%   \caption{Simulator Verification Result (time: ms)}
%   \label{tab:sim_veri}
%   \begin{tabular}{ccl}
%     \toprule
%     Model & Timeline & Simulator & Memcpy overhead & Gap between op & other overhead\\
%     \midrule
%     vgg_19 & 203.378 & 199.660 & 1.561 & 1.557 & 0.600\\
%     ResNet_50 & 282.557 & 277.470 & 1.567 & 2.634 & 0.886\\
%     ResNet_152 & 638.811 & 629.312 & 1.589 & 7.104 & 0.806\\
%   \bottomrule
% \end{tabular}
% \end{table}

Table~\ref{tab:simVeri} shows that our dataflow based simulation is accurate
with $<$2\% errors. The errors mainly come from memory copy 
and the time gap between ops that we have not considered.   
%!TEX root = main.tex
\section{Future Work and Conclusion}

We advocated offline profiling based simulation for ML systems in this work. To
fully demonstrate its feasibility and potential, we are conducting extensive
offline profiling to cover all operations in popular frameworks and common
hardware, and investigating the ML approach for op-level
performance estimation. We will also validate the effectiveness of our approach
on distributed settings with multiple machines and various networking
technologies.
We aim to develop a fully automated simulator together with the profiling
database and open source it for the community.
\begin{comment}
To fully  of our approach, 
Our current works have proved the feasibility of our key idea. For next step,
there are still some problems to solve:

{\bf Profiling Coverage.} Current profiling module cannot coverage profiling and modeling for all operations. We will continue upgrading the profiling module to support more operations and frameworks.

{\bf Op Performance Auto Modeling} To build a more efficient and accurate data base, we will try to upgrade the estimator feature with different techniques including regression, nn, etc.

%{\bf Strategy Coverage.} We will going on to support more communication strategy, such %as new all-reduce algorithms and parameter server structure. 

%{\bf On-device Resource Sharing.} In current design, we assume each device will execute %operations in FIFO. While in real training, some operations may share one device %resource simultaneously. E.g. multiple communication operations share the bandwidth. We %will investigate system behaviors in such cases and simulate them. 

{\bf Large Scale Experiments.} We have only finished experiments in single machine, multi GPU platform. In next step, we will conduct experiments on larger cluster.
\end{comment}

\bibliographystyle{ACM-Reference-Format}
\bibliography{main}

\end{document}